\documentclass[sigconf]{acmart}

\usepackage{xspace}
\usepackage{xcolor}

\AtBeginDocument{%
  }

\copyrightyear{2026}
\acmYear{2026}
\setcopyright{cc}
\setcctype{by}
\acmConference[ICSE-FoSE]{2026 IEEE/ACM 48th International Conference on Software Engineering: Future of Software Engineering }{April 12--18, 2026}{Rio de Janeiro, Brazil}
\acmBooktitle{2026 IEEE/ACM 48th International Conference on Software Engineering: Future of Software Engineering (ICSE-FoSE), April 12--18, 2026, Rio de Janeiro, Brazil}
\acmPrice{}
\acmDOI{10.1145/3793657.3793888}
\acmISBN{979-8-4007-2479-4/2026/04}

\begin{document}

\title{From Papers to Progress: Rethinking Knowledge Accumulation in Software Engineering}



\author{Jason Cusati}
\affiliation{%
  \institution{Virginia Tech}
  \city{Blacksburg}
  \state{VA}
  \country{USA}
}
\email{djjay@vt.edu}

\author{Chris Brown}
\affiliation{%
  \institution{Virginia Tech}
  \city{Blacksburg}
  \state{VA}
  \country{USA}
}
\email{dcbrown@vt.edu}

\renewcommand{\shortauthors}{Cusati and Brown}

\begin{abstract}
Software engineering research has experienced rapid growth in both output and participation over the past decades. Yet concerns persist about the field's ability to accumulate, integrate, and reuse knowledge in ways that support long-term progress. To better understand how the community itself perceives these challenges, we analyze responses from the ICSE 2026 Future of Software Engineering pre-survey, which captures perspectives from 280 globally distributed and highly experienced researchers. Our analysis reveals a tension between increasing research productivity and the limited mechanisms available for synthesizing results, tracking evolving claims, and supporting cumulative understanding over time.

Building on these observations, we diagnose four interrelated structural breakdowns: papers function as isolated knowledge units with claims embedded in prose; context and provenance are lost as knowledge moves through the publication pipeline; claims evolve without systematic tracking; and incentive structures favor novelty over consolidation. We argue that addressing these barriers requires rethinking the fundamental properties of research artifacts.

We articulate four technology-agnostic principles for future research artifacts: structured and interpretable representations of claims and evidence; inspectable and provenance-aware documentation of methodological decisions; long-lived and reusable substrates that evolve beyond publication; and governance mechanisms that align individual incentives with collective knowledge-building goals. We discuss implications for research practice, publication norms, and community infrastructure, positioning FOSE as a venue for experimenting with alternative artifact designs that support cumulative scientific progress.
\end{abstract}




\maketitle
\newcommand{\todo}[1]{{\color{red}\bfseries [[TODO: #1]]}}

\newcommand{\etal}{et al.\xspace}

\newcommand{\ie}{\textit{i.e.,}\xspace}

\newcommand{\eg}{\textit{e.g.,}\xspace}
\section{Introduction}
\label{sec:introduction}

Software engineering research is thriving by most measures. Conference submissions keep rising, publication venues have multiplied, and the community has gone global. Researchers are productive, techniques keep advancing, and new areas of inquiry emerge every year. But a persistent challenge lurks beneath all this activity: we struggle to build cumulative knowledge that actually connects, consolidates, and extends prior work.

This is not a new concern, nor one unique to our field. Across science, more publications have not automatically meant deeper understanding~\cite{ioannidis2005,fortunato2018}. In software engineering specifically, fragmentation and the difficulty of synthesizing results have been raised repeatedly~\cite{cruzes2010synthesizing,shull2008,sjberg2007}. Yet the structural factors limiting cumulative progress remain underexplored.

This paper offers a community-informed perspective on knowledge accumulation in software engineering. Drawing on 280 ICSE 2026 FOSE pre-survey responses from an experienced (57\% with 10+ years), productive (44\% with 11+ papers in 3 years), globally distributed community, we diagnose four structural breakdowns: papers as isolated units, lost context and provenance, untracked claim evolution, and incentives favoring novelty over consolidation. We then propose four principles for research artifacts: structured and interpretable, inspectable and provenance-aware, long-lived and reusable, and community-governed. We do not prescribe specific tools or claim to solve these challenges; instead, we synthesize community observations into a diagnosis and principles to guide future work.

\section{Community Signals from the ICSE 2026 FOSE Pre-Survey}
\label{sec:survey}

We ground our diagnosis in the ICSE 2026 Future of Software Engineering pre-survey (n=280)~\cite{storey_2025_18217799}. Two researchers performed a lightweight qualitative analysis, examining comments for insights about knowledge-building in software engineering. The respondents span six geographic regions (Europe 50\%, North America 25\%, Asia 15\%, others 10\%) and represent a mature, productive community. This profile matters for our argument: if knowledge accumulation remains limited despite all this expertise and output, the barriers are \textit{structural}, not individual. The community is not lacking effort. Respondents themselves recognize systemic challenges, noting fragmentation across resubmissions, difficulties building on prior work, and incentives that reward novelty over consolidation. These observations motivate our diagnosis.

\section{Where Knowledge Accumulation Breaks Down}
\label{sec:breakdown}

If the community is experienced, productive, and global, yet still struggles to synthesize results and build on prior work, the barriers are structural. We identify four interrelated breakdowns in how research artifacts are produced, represented, and connected. Anonymous participant identifiers appear in parentheses.

\subsection{Papers as Isolated Knowledge Units}

Research papers package claims, evidence, and context into narrative prose optimized for dissemination. This works well for presenting new ideas but creates barriers to cumulative building. One participant noted that ``\textit{15 years ago [it was the norm to] present and discuss a research idea at a workshop then go back home, build and evaluate the thing, then publish at a conference. These days, reviewers repeatedly reject such submissions `because the work has already appeared at this and that workshop'}'' (195811076). Claims get buried in paragraphs alongside motivation and methods. Extracting and comparing claims across papers means reading entire documents and reconciling different terminology. Evidence linkages stay implicit; readers infer which results support which claims under what assumptions. Even when artifacts are shared, results freeze at publication time. Updates to datasets or baselines require new papers, while original claims sit unchanged. As another participant observed, ``\textit{despite having a software artifact track, not many of the software artifacts are being actively maintained}'' (196030130). Each paper becomes an island connected only through citations and prose, requiring manual synthesis that does not scale.

\subsection{Loss of Context and Provenance}

Research involves countless contextual decisions: which dataset, how to split data, which baselines, how to handle edge cases. Papers describe these choices but often omit the rationale. One participant noted frustration with ``\textit{software engineering researchers trying to do research with generative AI tools without specifying context first}'' (196032564). As work gets cited and summarized, motivation and assumptions fade. Carefully qualified findings become unqualified facts in subsequent literature. Methodological decisions must be reverse-engineered, sometimes revealing that subtle choices significantly impacted results. Papers present polished final versions, hiding the path from hypothesis to result. When context and provenance are lost, later researchers either accept prior work at face value or invest substantial effort reconstructing the reasoning. Moreover, despite goals to enhance practice, research has ``\textit{minimal}'' and ``\textit{almost no relevance}'' to industry contexts (192101946).

\subsection{Claims Evolve Without Tracking}

Scientific claims get refined, qualified, contradicted, and superseded as evidence accumulates, but the publication system provides few tracking mechanisms. The literature may contain conflicting claims (technique A outperforms B in one paper, the reverse in another) without systematic flagging. Survey respondents noted that ``\textit{most of the SE research being published is neither reproducible nor replicable}'' (195853386) and ``\textit{we are reluctant to accept papers that replicate previous study...see medical sciences as counter example - they publish single cases as they allow to build body of konowledge in the long run}'' (196067582). When researchers discover conflicts, they must investigate causes themselves, often finding subtle methodological differences explain the divergence. Refinements stay implicit: a follow-on paper may qualify a prior claim, but the original remains unchanged. Claim relationships (does one extend, contradict, or depend on another?) appear only in natural language like "building on [12]" or "in contrast to [34]." Without structured representations, determining what is currently believed, under what conditions, and with what confidence requires extensive manual synthesis.

\subsection{Incentive Structures Favor Novelty Over Accumulation}

The final breakdown is embedded in incentive structures. Publication venues, hiring criteria, and funding mechanisms reward novelty~\cite{nosek2012,smaldino2016}. Conference and journal reviews prioritize new techniques; replication studies, negative results, and syntheses face higher acceptance bars even when they would advance collective understanding~\cite{basili1999}. Researchers investing in replication or shared infrastructure face opportunity costs: these efforts may not yield top-venue publications or count in promotion reviews~\cite{collberg2016}. One participant stated ``\textit{we are not good at replication (despite better rigour in what we do)...too much effort goes into chasing `new' rather than consolidating our knowledge}'' (192758693). Building knowledge repositories and interoperability layers requires sustained effort, often led by small groups without commensurate recognition. Everyone benefits from better infrastructure, but individuals face little incentive to contribute. As long as novelty is privileged over accumulation, knowledge stays fragmented. \\

These four breakdowns reinforce one another. Incremental fixes (better citation practices, more replications, improved repositories) help but are not enough. Addressing these barriers requires rethinking the fundamental properties of research artifacts themselves.

\section{Rethinking Research Artifacts for Cumulative Progress}
\label{sec:principles}

The structural barriers trace back to design choices in how research artifacts are produced and shared. By \textit{research artifacts} we mean the tangible outputs of research: papers, datasets, code repositories, benchmarks, computational notebooks, knowledge graphs, and experimental protocols. These artifacts are vessels for knowledge. They encode claims, evidence, methodological context, and the provenance of results. How well knowledge accumulates depends on their properties: whether claims can be extracted, evidence traced, artifacts reused beyond publication, and communities can build on prior work.

Addressing these barriers means reconsidering artifact properties. We propose four technology-agnostic principles. Implementations will vary, but these principles provide a shared foundation for evaluating research infrastructure.

\subsection{Principle 1: Structured and Interpretable}

\textbf{Principle:} Research artifacts should make claims, evidence, and context explicit and directly accessible, not just embedded in prose.

The first breakdown, papers as isolated units, stems from claims and evidence being woven into narrative text. Prose works well for explaining motivation, but it is not ideal for cumulative synthesis. Comparing claims across papers means reading entire documents and reconciling terminology. This does not scale as the literature grows.

Artifacts should represent claims, evidence, and context as structured, first-class entities. A claim should be identifiable: ``Technique X improves metric Y on dataset Z by $\delta$ under conditions C.'' Evidence and context should be explicitly linked, not buried in prose. Structured representations let conflicting claims become visible and resolvable by examining evidence directly. Claims referencing updated datasets can be systematically reevaluated. Examples include semantic annotations (machine-readable metadata describing claims and results) or knowledge graphs~\cite{orkg,wang2020,karras2024orkg,wang2023kg} (entities with typed relationships enabling queries like ``Which papers claim improvements on dataset D?''). Structure complements prose, making knowledge directly accessible while preserving narrative's explanatory power.

\subsection{Principle 2: Inspectable and Provenance-Aware}

\textbf{Principle:} Research artifacts should preserve the full provenance of claims, from raw data through methodological decisions to final results, and make this provenance inspectable (accessible and understandable by researchers who want to examine, verify, or build on the work).

The second breakdown, loss of context and provenance, occurs because reasoning behind decisions fades as knowledge moves through publication. Papers report final results but often omit the path taken: why this dataset, baseline, or protocol. Later researchers must reconstruct this reasoning, often discovering that subtle choices significantly impacted results.

Artifacts should document not just what was found but how and why. Every claim should trace to its sources: data, code, configuration, assumptions. Methodological decisions should be explicitly justified. When results are updated, the provenance chain should track changes~\cite{gonzalez-barahona2012}. Provenance enables trust: researchers can inspect a claim's lineage, verify evidence, and understand conditions. When claims conflict, provenance helps diagnose divergence. Examples include versioned computational artifacts (code and data in version control with clear lineage) or provenance graphs (explicit representations linking claims to experiments to data).

\subsection{Principle 3: Long-Lived and Reusable}

\textbf{Principle:} Research artifacts should support evolution and reuse, not remain static at publication.

Papers are snapshots frozen in time. Once published, they rarely update when new evidence emerges, datasets are revised, or methods improve. Follow-on work cites and describes differences in prose, but original claims remain unchanged. Research artifacts are also often hard to reuse. Prior efforts to replicate research systems from ICSE and FSE tool demonstration tracks succeeded in running only about half (76/131), despite thousands of hours of effort~\cite{murphyhill-reviving}.

Artifacts should be living substrates that can be updated, extended, and reused. When datasets are corrected, dependent claims should be re-evaluable. When techniques are refined, prior results should be comparable. This does not mean rewriting papers. Instead, underlying structured representations decouple from narrative documents. Narratives remain stable as historical records while structured substrates evolve. Examples include living knowledge bases (repositories where claims and datasets are versioned and updated) or executable benchmarks (evaluation frameworks rerun with updated data). Knowledge should outlive individual papers, accumulating in shared substrates.

\subsection{Principle 4: Governed with Human Oversight}

\textbf{Principle:} Research artifacts and infrastructures should be governed by community processes ensuring quality, integrity, and ethical responsibility.

Even with perfect technical infrastructure, cumulative progress requires coordination: quality standards, conflict resolution, and recognition for consolidation work. Artifacts cannot govern themselves. Structured representations and provenance tracking are valuable only if the community trusts, maintains, and uses them responsibly. This requires human oversight: peer review for knowledge contributions, curation, dispute resolution, and credit systems valuing infrastructure work alongside novel research. The human-artifact model suggests human expertise is necessary to understand artifacts within broader ecosystems and motivate future designs~\cite{b2011human}.

Governance also addresses ethical concerns: ownership, attribution, bias, and access. Who controls shared knowledge bases? How is credit assigned for incremental contributions? How do we prevent infrastructures from perpetuating biases? These questions require community deliberation and ongoing stewardship. Examples include community curation processes (peer-reviewed contributions with clear criteria) or credit systems recognizing infrastructure, replication, and consolidation work. Cumulative progress is a collective endeavor; technical solutions must be embedded in social practices that align individual incentives with collective goals.



\section{From Principles to Practice: Implications for the Future of Software Engineering}
\label{sec:implications}

These four principles provide a framework for evaluating incremental steps: does a proposed tool or practice make artifacts more structured, inspectable, reusable, or better governed? The principles are technology-agnostic. Knowledge graphs, computational notebooks, and repositories are possible instantiations, but the principles describe \textit{properties}, leaving room for diverse implementations. The challenge is designing systems and reforming incentives so cumulative knowledge building becomes both possible and rewarded.

\subsection{Research Practice and Publication}

Research practice would shift toward documenting process alongside outcomes: recording methodological decisions and their rationales using computational notebooks, version control, and workflow platforms. Researchers would supplement narrative papers with structured representations (semantic annotations or knowledge graph entries) making contributions directly accessible for synthesis. Designing for reuse would become standard: datasets documented with schemas, code modular and documented, protocols reproducible.

Publication and review would need to value consolidation alongside novelty. Papers that resolve contradictions, curate benchmarks, or provide infrastructure deserve recognition. These replications are critical to avoid contributing to the ``vast graveyard of undead theories'' in software research~\cite{ferguson2012vast}. Conferences should treat structured artifacts as first-class contributions, not optional supplements~\cite{liu2024artifacts}. As one participant noted, replication studies are considered ``\textit{thankless volunteer work}'' and ``\textit{it is not sustainable for people to be doing these things well and frequently without proper incentives}'' (196051220). Replication efforts in other domains found only about one-third of published experiments in psychology (39/100)~\cite{open2015estimating} and economics (22/67)~\cite{chang2015economics} could reproduce statistically significant results. If artifacts evolve post-publication, papers become snapshots while living artifacts accumulate evidence, requiring new norms for citing evolving work.

\subsection{Community Infrastructure}

The community needs platforms for storing, querying, and updating structured artifacts (knowledge graphs, benchmark repositories, collaborative platforms) maintained through community governance~\cite{atkins2003,tamasauskaite2023kgdev}. One survey participant proposed ``\textit{a shared repository for research ideas, datasets, and research materials that lead to better distribution, collaboration, and replication}'' (196181126). Shared standards (ontologies, schemas, reporting protocols) enable reuse across studies~\cite{wohlin2012}. Infrastructure contributions and replications need recognition through alternative metrics or contribution-tracking platforms~\cite{allen2019}. Governance structures manage stewardship: maintaining repositories, resolving disputes, addressing ethical concerns. Without such infrastructure, the principles stay aspirational.

\subsection{FOSE as a Venue for Experimentation}

The Future of Software Engineering track could support experimentation with alternative artifact types: submissions experimenting with structured artifacts or living documents, judged on potential to demonstrate new knowledge-organizing approaches rather than solely novelty.

Consider a study finding automated code review reduces defects by 15\%. Under current norms, this paper gets cited and superseded as its dataset ages. Under our principles, the group registers structured claims in a knowledge graph, links them to versioned data with provenance, and designs for reuse. When contradictions emerge, structured representations enable diagnosing whether differences stem from datasets or methods, turning disputes into refinement. FOSE could welcome such experiments, building evidence for broader adoption.

\section{Limitations}
\label{sec:limitations}

This paper diagnoses structural barriers and proposes principles, but does not provide complete solutions.

\textbf{Survey limitations.} Our analysis draws on 280 FOSE pre-survey responses, a self-selected fraction of the global community. The data are descriptive: they establish that the community is experienced and productive, supporting our argument that barriers are structural, but they do not prove specific breakdowns. The survey reflects perspectives at one moment, and our principles may not generalize to other domains.

\textbf{Feasibility challenges.} Producing structured, provenance-aware artifacts takes effort that researchers may lack without support. The needed infrastructure (shared repositories, standards, governance) does not yet exist in many areas~\cite{hermann2020}. Adoption depends on changing incentives at multiple levels, which is difficult and slow.

\textbf{Ethical concerns.} Structured knowledge infrastructures raise questions about ownership, bias, access, and privacy. These are not purely technical issues; they require community deliberation and ongoing stewardship.

\section{Conclusion}
\label{sec:conclusion}

Software engineering research is productive and globally distributed, yet knowledge accumulation lags behind knowledge production. Drawing on survey responses from the community, this paper argues the barriers are structural: papers function as isolated documents, provenance is lost, claims evolve without tracking, and incentives favor novelty over consolidation. We propose four principles to guide future work: structured and interpretable, inspectable and provenance-aware, long-lived and reusable, and governed with human oversight.

These principles are aspirational and technology-agnostic. Realizing them requires changes in practice, publication norms, infrastructure, and incentives. The challenges are significant, but continuing with practices optimized for dissemination rather than accumulation will perpetuate fragmentation as the field grows.

The Future of Software Engineering track provides space for this conversation. Next steps include experimenting with alternative artifact designs, developing aligned infrastructure, and revising recognition systems to value consolidation alongside novelty. The community must engage in dialogue about what cumulative progress requires and what trade-offs are acceptable.

The future of software engineering depends not only on what we discover but on how we organize and preserve what we know. Building infrastructures that support cumulative knowledge building, making it rewarded and not just possible, is essential to ensuring the next generation inherits not just a literature to read but a knowledge base to build upon.

\bibliographystyle{ACM-Reference-Format}
\bibliography{bibliography,software}



\end{document}